\documentclass[aps,prd,preprintnumbers,superscriptaddress,nofootinbib,amsmath,amssymb,showpacs]{revtex4}
\usepackage{graphicx}
\usepackage{dcolumn}
\usepackage{bm}
\usepackage{amsmath}
\usepackage{color}

\input{colordvi.tex}

\newcommand{\beq}{\begin{equation}}
\newcommand{\eeq}{\end{equation}}
\newcommand{\beqa}{\begin{eqnarray}}
\newcommand{\eeqa}{\end{eqnarray}}

\begin{document}

\begin{flushleft}
RESCEU-36/14 \\
\end{flushleft}

\title{Effects of cosmic strings with delayed scaling on CMB anisotropy}

\author{Kohei Kamada}
\email[Email: ]{kohei.kamada"at"epfl.ch}
\affiliation{ Institut de Th\'eorie des Ph\'enom\`enes Physiques,
\'Ecole Polytechnique F\'ed\'erale de Lausanne, 1015 Lausanne, Switzerland}

\author{Yuhei Miyamoto}
\email[Email: ]{miyamoto"at"resceu.s.u-tokyo.ac.jp}
\affiliation{ Department of Physics, Graduate School of Science, \\
The University of Tokyo, Tokyo 113-0033, Japan}
\affiliation{Research Center for the Early Universe (RESCEU), \\
Graduate School of Science, The University of Tokyo, Tokyo 113-0033, Japan}

\author{Daisuke Yamauchi}
\email[Email: ]{yamauchi"at"resceu.s.u-tokyo.ac.jp}
\affiliation{Research Center for the Early Universe (RESCEU), \\
Graduate School of Science, The University of Tokyo, Tokyo 113-0033, Japan}

\author{Jun'ichi Yokoyama}
\email[Email: ]{yokoyama"at"resceu.s.u-tokyo.ac.jp}
\affiliation{Research Center for the Early Universe (RESCEU), \\
Graduate School of Science, The University of Tokyo, Tokyo 113-0033, Japan}
\affiliation{Kavli Institute for the Physics and Mathematics of the Universe (Kavli IPMU),
WPI, TODIAS, The University of Tokyo, Kashiwa, Chiba, 277-8568, Japan}


\begin{abstract}

The network of cosmic strings generated in a phase transition during inflation 
enters the scaling regime later than that of usual strings.
If it occurs after the recombination, 
temperature anisotropies of the cosmic microwave background (CMB)
at high multipole moments are significantly reduced.
In this paper, we study such effects qualitatively and show that the constraint on
the cosmic string tension from the CMB temperature anisotropies and B-mode polarizations can be relaxed. 
It is shown to be difficult to explain the recent BICEP2
and POLARBEAR results in terms of signals induced by cosmic strings alone
even if we take into account the delayed scaling.
However, the inflationary tensor-to-scalar ratio 
required to explain the observed B-mode signals
can be slightly reduced to be consistent with the Planck constraint.

\end{abstract}
\maketitle

\section{Introduction}

Phase transitions are ubiquitous in the broad fields of physics and, in particular, the key ingredients of  
high energy physics such as grand unified theories.  
They predict formation of topological defects \cite{stringreveiw} through the Kibble mechanism \cite{Kibble:1976sj}
such as monopoles, domain walls, and 
cosmic strings \cite{cosmicstring}, depending on the symmetry-breaking pattern. 
Among various types of topological defects, cosmic strings are intriguing objects 
both in cosmology and high energy physics.
They may be produced abundantly in the early Universe
but do not dominate the subsequent Universe 
due to their scaling feature \cite{Kibble:1984hp}, and hence they can leave a variety of traces such as the
gravitational-wave background \cite{Vilenkin:1981bx}
or the cosmic microwave background (CMB) temperature/polarization anisotropies \cite{Albrecht:1997nt,Seljak:1997ii}
without overclosing the Universe, 
which gives us rich information about the high energy physics. 

In the standard scenario, cosmic strings are formed just after inflation 
or through thermal phase transition and enter the scaling regime well before the recombination. 
Therefore, their effects on the CMB temperature/polarization anisotropies are 
almost independent of the initial condition, 
which allows us to give general constraints on their properties, such as their tension.   
However, as discussed in the 1980s \cite{Lazarides:1984pq} and recently pointed out again in a modern context 
\cite{Kamada:2012ag}, 
cosmic strings can be formed {\it during} inflation and they can enter the scaling regime at a later epoch
since they are diluted partially during subsequent inflation. 
In this paper, we show the qualitative effects of this ``delayed scaling'' of cosmic strings on the CMB anisotropies. 

Observations of the CMB temperature anisotropies by WMAP \cite{Hinshaw:2012aka} and Planck 
\cite{Ade:2013uln} strongly 
support the simplest models of inflation, namely, the slow-roll canonical single-field inflation models, 
and it is found that the cosmic strings alone can explain neither the CMB temperature anisotropies 
nor the present large scale structure of the Universe. 
In turn, the CMB temperature anisotropies give a severe constraint on the cosmic string tension of
$G\mu \lesssim 1.3 \times 10^{-7}$ for the Nambu-Goto cosmic string model
and $G\mu \lesssim 3.2 \times 10^{-7}$ for the Abelian Higgs cosmic string model \cite{Ade:2013xla}. 

Recent observations of the B-mode signal of the CMB polarization anisotropies reported by BICEP2 
\cite{Ade:2014xna} 
also give notable suggestions of the physics of the early Universe and the high energy physics.  
Again, the observed B-mode signal at low multipoles can be explained by the inflationary gravitational waves 
with the tensor-to-scalar ratio, which directly relates to the energy scale of inflation, being around $r\simeq 0.2$.\footnote{
Many authors use the Planck constraint  on the tensor-to-scalar 
ratio $r<0.11$ at $k_0=0.002{\rm Mpc}^{-1}$ for 
$n_t=-r/8$~\cite{Ade:2013uln} to compare it with the BICEP2 result.
In Refs.~\cite{Audren:2014cea}, however, 
it has been shown that  at our pivot scale $k_0=0.05{\rm Mpc}^{-1}$ for $n_t=0$, 
the Planck data give the constraint as $r<0.135$.
Therefore we take $r=0.135$, $n_t=0$ throughout this paper when we show effects of inflationary tensorial perturbations.}
It is also found that cosmic strings alone cannot fully explain the B-mode signal, since 
even if we try to fit the BICEP2 B-mode data at the low multipoles by cosmic strings, 
the dip around $\ell \sim 150$ in the BICEP 2 data is difficult to fit fully and
it is incompatible with 
those at high multipoles observed by POLARBEAR \cite{Ade:2014afa}
and temperature anisotropies observed by Planck \cite{Lizarraga:2014eaa}. 

However, the above conclusion is based on the assumption that the cosmic string network enters the 
scaling regime well before the recombination. Therefore, it is worth studying how the discussion changes 
if we take into account the delayed scaling scenario of cosmic strings \cite{Lazarides:1984pq,Kamada:2012ag}. 
In this paper, we study its qualitative effect 
and show that it relaxes the constraint since 
the source term for the CMB temperature anisotropies is reduced. 
Moreover, we find that the B-mode signals may give a stronger constraint on the cosmic string tension, 
depending on the time when the system enters the scaling regime. 
Even if we take into account the delayed scaling, 
it is shown to be difficult to explain the BICEP2 and POLARBEAR B-mode
signals by the cosmic string alone, but we also exhibit that they can be fit well by the combination of the cosmic strings, primordial 
gravitational waves and the gravitational lensing without conflicting with other constraints from the Planck observation. 

This paper is organized as follows. 
In Sec.~\ref{sec:2}, we introduce the delayed scaling scenario using the velocity-dependent one-scale model \cite{Kibble:1984hp,Martins:1996jp,Martins:2000cs}.  
In Sec.~\ref{sec:3}, we discuss how it changes the observational signatures in the CMB qualitatively. 
We also discuss its impact on BICEP2 result. 
Section~\ref{sec:4} is devoted to  conclusion and discussion.
Throughout the paper, our fiducial model is the standard $\Lambda$CDM cosmological model
with the cosmological parameters $\Omega_{\rm b}h^2=0.022161$, $\Omega_{\rm m}h^2=0.11889$\,, and
$h=0.6777$ with $\Delta_{\mathcal R}^2 =2.216\times 10^{-9}$ and  $n_{\rm s}=0.96$, at $k_0=0.05{\rm Mpc}^{-1}$,
which is the best-fit model of the Planck 2013 data combined with the polarization of 
WMAP, other high-$\ell$ observations, and the baryon acoustic oscillation \cite{Ade:2013zuv}.

\section{Delayed scaling scenario \label{sec:2}}

Let us first summarize the delayed scaling scenario. 
Symmetries are naturally restored during inflation due to 
the ``Hubble-induced'' mass for the Higgs field responsible for the symmetry breaking\footnote{For the 
similar studies on the monopole production during inflation, see Ref.~\cite{Yokoyama:1989xj}.} 
coming from, for example, 
the direct coupling to the inflaton,  the nonminimal coupling to Ricci scalar \cite{Lazarides:1984pq}, 
or the Planck suppressed interaction in the  F-term inflation \cite{Kamada:2012ag}. 
If the induced mass is larger than the amplitude of its tachyonic zero-temperature mass, 
the Higgs field settles down to the origin during inflation and the symmetry is restored. 
In the case the induced mass decreases with time significantly enough, during inflation
it can become smaller than the amplitude of the zero-temperature mass 
and cosmic strings can be formed thorough the phase transition. 
In this case, they are diluted and their correlation length and mean separation expand exponentially 
until the end of inflation. 
As a result, their average distances become much larger than the Hubble length
at the end of inflation.  
This large separation allows the cosmic string network to enter the scaling regime, 
where the characteristic scale of the string network remains constant relative to the 
Hubble length, at a later time. 
This is because the scaling regime is achieved after the characteristic length of the string network 
becomes smaller than the Hubble length and the energy of infinite strings starts to 
be converted into small string loops. 

In principle, numerical simulations are necessary to study how the system enters 
the scaling regime in this delayed scaling scenario. 
Since it will need a huge computation, here we instead evaluate the evolution of the cosmic string 
network using the one-scale model \cite{Kibble:1984hp,Martins:1996jp,Martins:2000cs}
as a first step. 
In the (velocity-dependent) one-scale model \cite{Martins:1996jp,Martins:2000cs}, 
the correlation length of the infinite string $L$, 
which represents both the typical curvature radius of infinite strings and
their mean separation,
obeys the evolution equation
\begin{equation}
\label{voseq1}
\frac{dL}{dt}=(1+v^2) H L+\frac{1}{2}{\tilde c} v\,.
\end{equation}
Here $t$ is the physical time, $H={\dot a}/a=H_0(\Omega_{\rm m} a^{-3}+\Omega_{\rm r} a^{-4} +\Omega_\Lambda)^{1/2}$ 
is the Hubble parameter, $v$ is the 
root mean square of the velocity of infinite string, and ${\tilde c}=0.23$ is the numerical 
parameter that represents the loop formation efficiency \cite{Martins:2000cs}. 
We set the scale factor $a=1$ today. $H_0$ is the present Hubble parameter, and 
$\Omega_{\rm m}, \Omega_{\rm r}$, and $\Omega_\Lambda$ are the present density parameters of the 
nonrelativistic  matter, relativistic matter, and the cosmological constant, respectively. 
The velocity $v$ evolves with the evolution equation, 
\begin{equation}
\label{voseq2}
\frac{dv}{dt}=(1-v^2)\left(\frac{\tilde k}{L}-2 Hv\right), 
\end{equation}
with ${\tilde k}=(2\sqrt{2}/\pi)((1-8v^6)/(1+8v^6))$ being the momentum parameter that represents 
the acceleration effect due to the curvature of the strings \cite{Martins:2000cs}.

Although the velocity-dependent one-scale model, characterized by Eqs.~(\ref{voseq1}) and (\ref{voseq2}),
is intended to describe evolution of the string network formed by the conventional Kibble mechanism,
these equations also reproduce the initial evolution of string segments 
before entering the scaling regime correctly; 
that is, $v$ tends to 0 when $L\gg H^{-1}$ and $L$ evolves in proportion to the scale factor, 
if we take an appropriate ``initial'' time with a large initial correlation length $L_{\rm ini}\gg H_{\rm ini}^{-1}$. 
Note that the CMB anisotropies induced by cosmic strings are insensitive to their behaviors in the earlier epoch, 
and we do not have to follow their evolution from the end of inflation. 
It is not clear if the one-scale model, where we assume that the typical curvature of the infinite strings 
and their mean separation are equal, holds just after inflation, but we expect it gives a good approximation 
since the Hubble parameter during inflation is the unique parameter to determine the string configuration 
when they are formed.
The validity of this model, especially in the intermediate regime, 
should nevertheless be investigated through numerical simulations with appropriate initial conditions,
which is beyond the scope of the present paper, and we use (\ref{voseq1}) and (\ref{voseq2}) throughout.

\begin{figure}[htbp]
\centering{
\includegraphics[width = 0.45\textwidth]{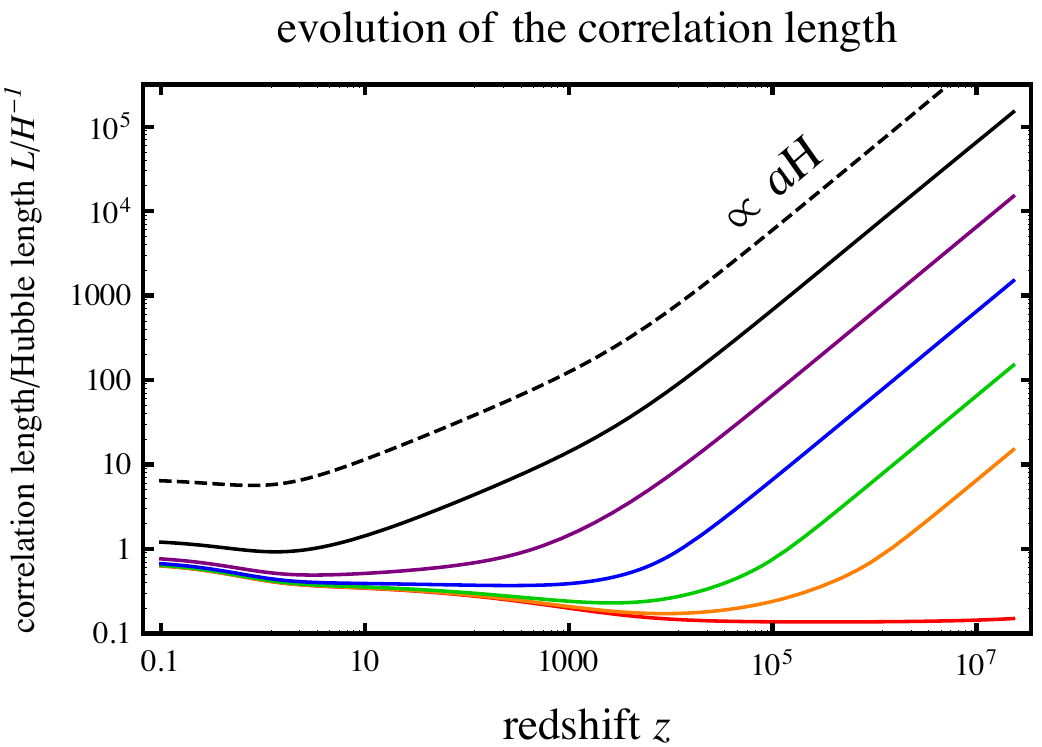}}
\caption{The evolution of the correlation length relative to the Hubble length.
 Compared to the dashed black line which is proportional to $aH$,
 we can easily see that the correlation length evolves proportional to the scale factor on superhorizon scales. }\label{fig1}
\end{figure}

\begin{figure}[htbp]
\centering{
\includegraphics[width = 0.45\textwidth]{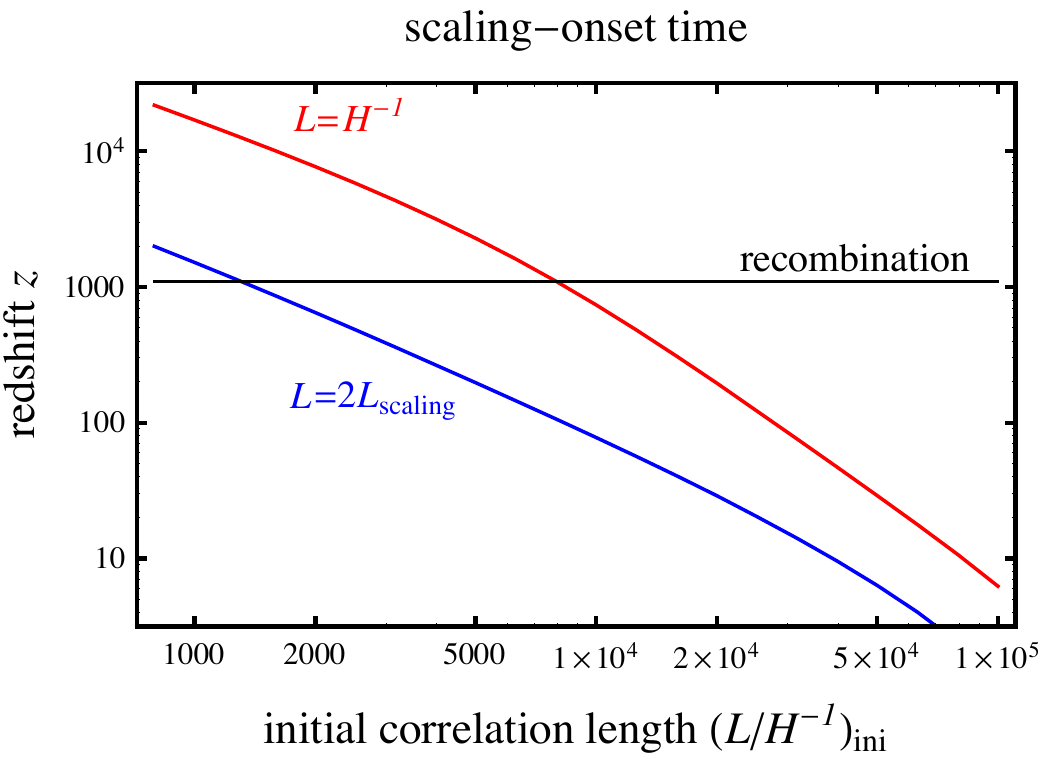}}
\caption{The redshift at which the system enters the horizon and approaches
the scaling solution as a function of the initial correlation length.}
\label{fig:zsc}
\end{figure}

Figure~\ref{fig1} shows the typical evolution of the correlation length relative to the 
Hubble length $H^{-1}$ with a different (relatively large) initial correlation length. 
Hereafter we set $z=2.3 \times 10^7$ as initial time.
The initial velocity is set to $v=0$ except for the bottom line, 
which represents the standard, always-scaling case with initial velocity $v=0.65$.
While we chose such initial velocities, 
we also confirmed that the evolution of the correlation length is almost 
independent of the initial velocity, since the velocity 
decreases vanishingly  and it loses its initial information quickly. 
As mentioned above, when $L$ is larger than the horizon scale,
it simply evolves in proportion to $a$.
In terms of the redshift $z$,  $L/H^{-1}$ is proportional to $z$ in the radiation 
dominated era ($z\gg z_{\rm eq}$) and $z^{1/2}$ in the matter dominated era ($z \ll z_{\rm eq}$),
where $z_{\rm eq}\approx 3400$ is the redshift at the matter-radiation equality. 
We can also see that it takes a few orders of redshift for the system 
to enter the scaling solution completely, which will be important for the observational signatures.
Since this result shows the general feature of the evolution of the correlation length on superhorizon scales, 
it strongly suggests that even if the correlation length is much larger than the Hubble length just after inflation,
the system gradually approaches the scaling solution and 
at a relatively late epoch, say $z=10^3$  or later,
starts to evolve in accordance with the scaling rule
depending on the epoch of the phase transition during inflation.
Note that if the phase transition takes place when the present horizon scale exited the horizon 
during inflation, 
the correlation length would become the horizon scale again today, 
since its initial correlation length can be estimated by the horizon scale at that time. 
Therefore, cosmic strings formed several  $e$-folds after the current Hubble scale went out of the horizon during inflation
would enter the scaling regime after the recombination.\footnote{In principle, the initial correlation length 
can be calculated from the model parameters, (see, {\it e.g.}, Ref.~\cite{Kamada:2012ag}), 
but the inflation model must be specified, which is beyond the scope of this study. }
To evaluate the onset time of scaling,
we calculate the redshift when the correlation length $L$
becomes $2L_{\rm scaling}$, as well as the redshift when the correlation length falls shorter than the horizon.
Here $L_{\rm scaling}$ is the correlation length of the scaling network, which corresponds to the bottom line in Fig.~\ref{fig1}.
In Fig.~\ref{fig:zsc}, we show these redshifts as a function of the initial correlation length.
This figure shows that there arises an intermediate epoch before the complete entrance into the scaling regime, 
which may be important for the CMB anisotropies.

\section{CMB from cosmic strings with delayed scaling \label{sec:3}}

We now study the signatures of cosmic strings on the CMB anisotropies
in the delayed scaling scenario. 
Here we use a code based on the public code CMBACT \cite{Pogosian:1999np}.
For comparison, we calculate the inflationary contributions using CAMB \cite{Lewis:1999bs}.

\subsection{Temperature anisotropies}

\begin{figure*}[tbp]
\centering{
\includegraphics[width = 1.0\textwidth]{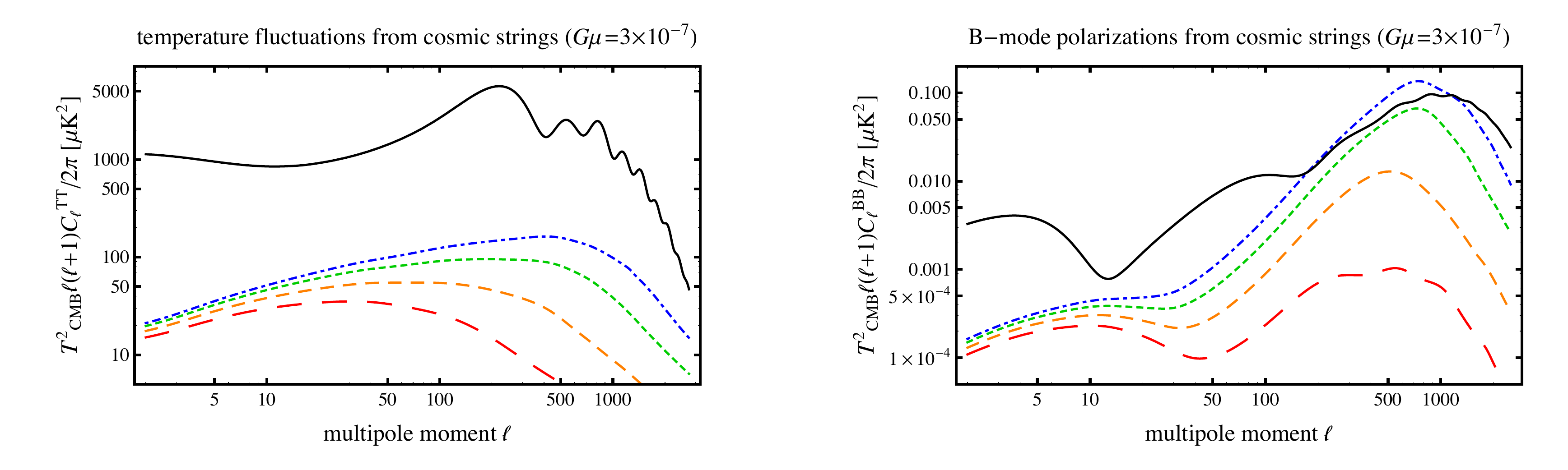}}
\caption{Angular power spectra for the temperature/B-mode polarization fluctuations
 induced by cosmic strings with $G\mu=3\times10^{-7}.$
 From top (blue dot-dashed) to bottom (red long dashed),
we take the initial correlation length $(L/H^{-1})_{\rm ini} = 1.5,
7.5 \times 10^3, 4.5 \times 10^4$, and $1.5 \times 10^5$. 
In the left panel, the contributions from
inflationary perturbations are shown in the black solid line, for comparison.
In the right panel, the black solid line corresponds to the primordial gravitational waves with
$r$ = 0.135 and gravitational lensing.
Note that the overall amplitudes scale as $(G\mu)^2$.}
\label{fig:TTandBB}
\end{figure*}

Let us discuss first the effect of  strings with delayed scaling on 
the temperature anisotropies to investigate how the constraint on the string tension changes
qualitatively.
Varying the initial correlation length, we show the angular power spectrum 
for the CMB temperature anisotropies induced by cosmic strings in the left panel in Fig.~\ref{fig:TTandBB}.
For illustrative purposes, from top to bottom, we adopt the string tension 
$G\mu =3\times 10^{-7}$ with the initial correlation lengths as 
$(L/H^{-1})_{\rm ini}=1.5$, $7.5\times 10^3$, $4.5\times 10^4$, and $1.5\times 10^5$.
With the help of Fig.~\ref{fig:zsc},
we can estimate the scaling-onset time for each initial correlation length.
Hence, we can treat the last three cases as the delayed scaling cases,
while the first one corresponds to the system which already realizes the scaling
by the time of the recombination.

One can see that the contributions from the 
strings give a single broad peak.
It moves to the lower multipoles and 
the power-law behavior on small scales decays
as the initial correlation length becomes large.
Thus, increasing $(L/H^{-1})_{\rm ini}$ decreases the string-induced small-scale signals significantly.
On large-scale signals, $\ell \sim10$, 
we also see a slight decrease of the amplitude of the temperature anisotropies.

Now we interpret the above features of the strings with delayed scaling.
In considering the effect of the long string network on the CMB, 
we can treat it as if it consisted of many ``string segments'' with length $L$ \cite{Pogosian:1999np}.
Since in the one-scale model the number density of such segments of cosmic strings is estimated as $L^{-3}$,
the number density and the resultant contributions to the CMB fluctuations are small when the correlation length is large.
With this fact in our mind, 
we can presume the reason of the significant decrease at high multipoles as follows.
In the standard scenario, the peak around 
$\ell \sim 500$ is generated by strings at the recombination \cite{Bevis:2010gj}.
Therefore, 
when the correlation length of strings is very large at the recombination,
the number density of strings and their contributions to such high multipoles become significantly small.
In this case, since the number density of strings is negligibly small 
until the system enters the scaling regime,
the position of the peak would be determined by the onset time of scaling and get lower.
Next, we consider why we see only a slight decline at low multipoles.
The signal in these scales would be mainly induced by the cosmic strings at late times.
As we can see from Figs.~\ref{fig1} and \ref{fig:zsc}, it takes time for the system to enter the complete scaling regime 
and a larger initial correlation length turns into a slightly larger correlation length
and hence slightly smaller number density at later times. 
As a result, there is only a slight decrease of the amplitude of 
the string-induced large-scale signals.\footnote{
Note that while the number of strings is reduced, 
each segment contributes to larger multipoles,
since we expect the scale of dominant fluctuations generated by each segment to be proportional to $L^{-1}$.
Therefore the delayed scaling effect is not fully determined by the number density.
}

\begin{figure}[htbp]
\centering{
\includegraphics[width = 0.45\textwidth]{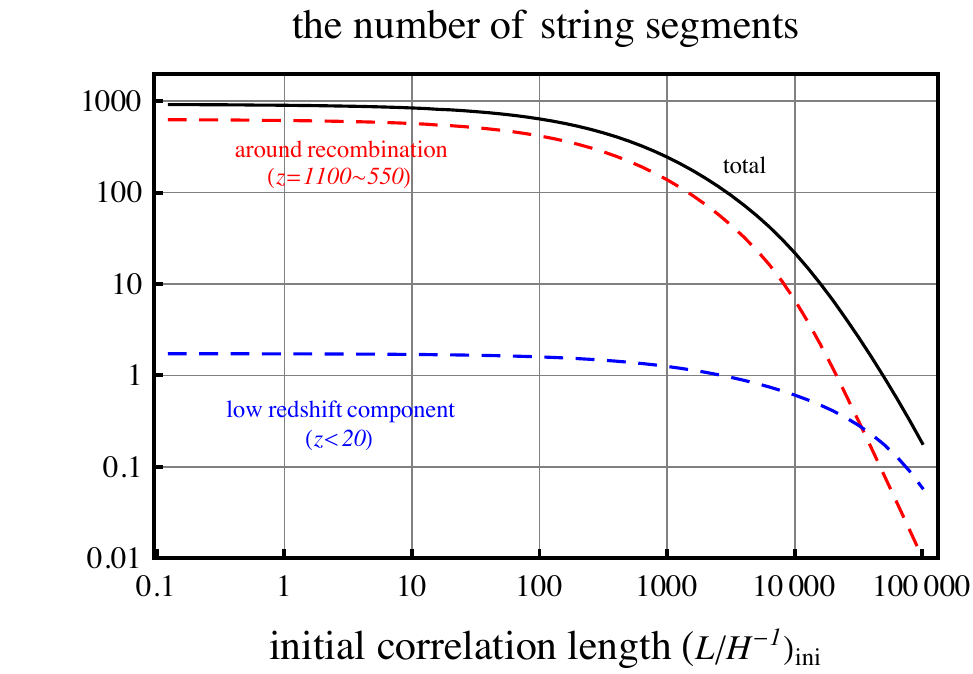}}
\caption{
The solid black line represents the total number of string segments from the recombination to the present.
The dashed red line, which corresponds to the number of strings near the last scattering surface, 
rapidly decreases as the initial correlation length becomes large.
We also plot the number of strings in the low-redshift region ($0\leq z\leq 20$) by a dashed blue line.
Its dependence on the initial correlation length is smaller than that of strings being around the recombination.
}
\label{fig:total number}
\end{figure}

Here we comment more on the number of cosmic string segments. 
We define the total number of cosmic strings between the last scattering surface and us as 
\begin{align}
&\int^{r_{\rm dec}}_0 4\pi a^3 r^2(z)dr \frac{1}{L^3(z)} \notag\\
=&\int_0^{1100} \,\frac{dz}{H(z)} \frac{4\pi}{(1+z)^3} \left( \int _0^{z} \frac{1}{H(z')}dz' \right)^2 \frac{1}{L^3(z)}\,,
\end{align}
where $r(z)$ is a comoving distance to the surface whose redshift is $z$,
\begin{equation}
r(z)=\int_0^z \frac{1}{H(z')}dz'.
\end{equation}
In Fig.~\ref{fig:total number},
we show the dependences on the initial correlation length
for the total number and its partial components which are expected to give dominant contributions to large- and small-scale fluctuations. 
We can explicitly see that the total number of cosmic strings is
significantly reduced if we set the initial correlation length very large.
We can also see that for $(L/H^{-1})_{\rm ini}=10^3 \sim 10^5$, the number of string segments around the recombination 
drastically decreases, whereas those around reionization show a milder decrease. 
This is consistent with
the behavior of the power spectrum of the CMB temperature and polarization fluctuations in Fig.~\ref{fig:TTandBB}.
Note that for the large initial correlation length, the cosmic variance becomes so large that 
we have to be careful in comparing theoretical, ensemble-averaged quantities and observations.

With the discussion given above, we then conclude that 
the constraint on the string tension is relaxed in the delayed scaling scenario.
\begin{figure*}[htbp]
\centering{
\includegraphics[width =1.0\textwidth]{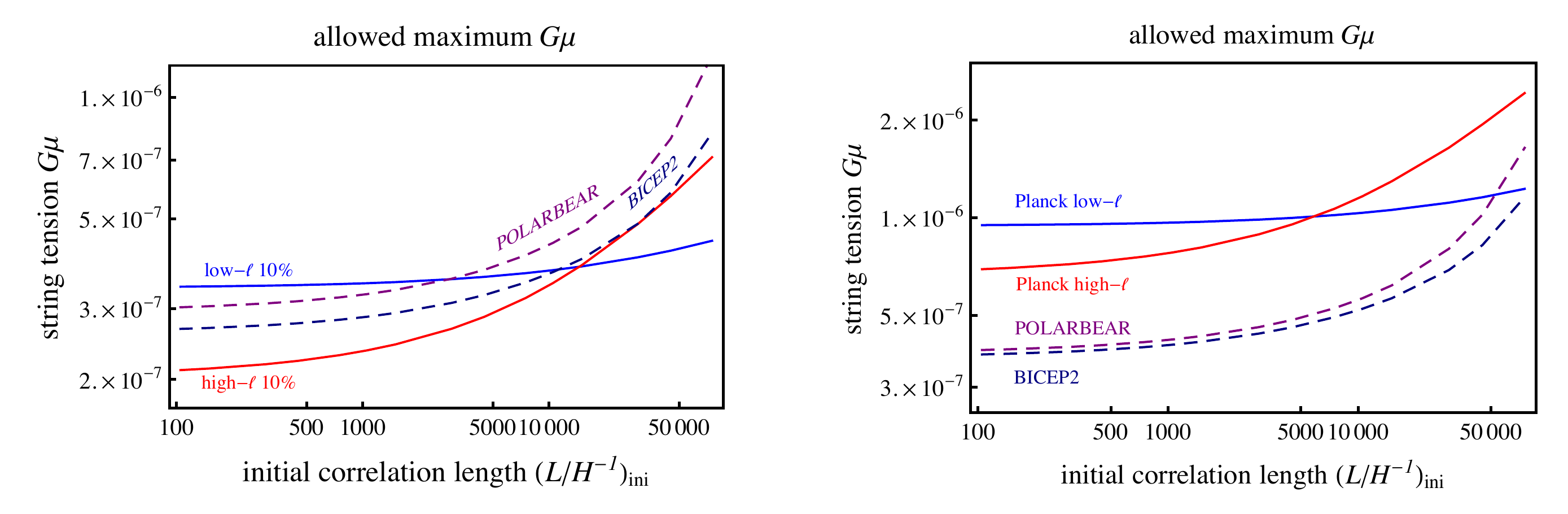}}
\caption{
The contour for the constraint on the string tension
$G\mu$ as a function of the initial correlation length $(L/H^{-1})_{\rm ini}$.
In the left panel, constraints obtained from the condition that the
string temperature anisotropies do not exceed $10\%$ of the primordial
one are shown in red (small scale, $2250 \leq \ell \leq 2450$) and blue (large scale, $\ell \leq 50$). 
The dashed dark blue (dark red) line shows the excluded region
from the condition that the string polarizations added to the lensing effect do not
exceed the spectra measured by the BICEP2 (POLARBEAR) data. 
In the right panel, more conservative limits on $G\mu$ are depicted from  
the condition that the temperature and polarization fluctuations generated only by cosmic strings
(namely, without inflationary temperature fluctuations and gravitational waves, and the lensing effect) 
do not exceed the observed values.
}
\label{fig:gmulimit}
\end{figure*}
We show some quantitative constraints in Fig.~\ref{fig:gmulimit}.  
Solid lines in the left panel of Fig.~\ref{fig:gmulimit} show the upper bound on the string tension from the CMB
temperature anisotropies,
which was obtained from the condition that the string-induced temperature
anisotropies would not exceed $10\%$\footnote{
The precision of the data provided by Planck is about 10\%  including cosmic variance
for $\ell \lesssim 50$ and $2250 \lesssim \ell \lesssim 2450$. }
 of the values given by the fiducial $\Lambda$CDM model.
The red line is the constraint from the small-scale signals, whereas the blue line is that from the large-scale counterparts. 
We can see that the small-scale signals give a stronger constraint for smaller initial correlation length and large-scale signals give 
a stronger one for larger initial correlation length. 
In particular, for $(L/H^{-1})_{\rm ini}=10^4$, the constraint on $G\mu$ is $G\mu\lesssim 3.4\times 10^{-7}$.
Although the resultant constraint would depend on the criterion of the condition, 
the generic features are expected to remain the same.
For comparison, in the right panel of Fig.~\ref{fig:gmulimit} 
we also show more conservative limits coming from the condition that
the temperature fluctuation created solely by cosmic strings (without inflationary perturbations) would not exceed 
the observed value by Planck.
We can see that although the quantitative constraints are very different,
shapes of the constraint lines do not change significantly.
That is, the constraint from the small scales is more severe than that from the large scales for smaller initial correlation length, 
and the opposite is the case for longer initial correlation length.

In summary, we emphasize that 
the high-$\ell$ temperature power spectra from cosmic strings are suppressed 
due to the delayed scaling, which loosens the constraint on the string tension. 
Note that the present constraint on the string tension comes from high-$\ell$ data.

\subsection{B-mode polarizations}

Let us discuss the effect of the delayed scaling on the CMB polarization signals.
The right panel of Fig.~\ref{fig:TTandBB} shows the angular power spectra for the B-mode polarization 
from strings with delayed scaling with $G\mu =3\times 10^{-7}$ and the initial correlation lengths
$(L/H^{-1})_{\rm ini}=1.5$, $7.5\times 10^3$, $4.5\times 10^4$ and $1.5\times 10^5$.
The resultant spectrum has two broad peaks, whose positions are expected to 
be determined by the contributions from the scattering of photons during the reionization 
(the peak at lower multipole) and the recombination epochs (the peak at higher multipole).
We can see that the main peak at higher multipole 
rather than the peak at lower multipole is sensitive to the initial correlation length.
Relatively large initial correlation length in general leads to the decrease in the B-mode signals on small scales.
This is mainly because increasing $(L/H^{-1})_{\rm ini}$ leads to 
the decrease in the number of strings during the recombination. 
We also see that the scale of the peak does not shift so much according to the change of the initial correlation length. 
If the position of the peak is determined only by the correlation length $L$ at the recombination, 
we expect it would be located at much lower multipole. 
The reason for this not being the case would be the fact that
we do not observe the snapshot
of the string network itself at the recombination which does not occur
instantaneously.
In contrast, the peak at lower multipole is not suppressed drastically.
This would be understood by the same discussion in the previous subsection.
From Fig.~\ref{fig:total number}, we can see that the number of strings at the reionization is less dependent on
the initial correlation length than those at the recombination for the initial correlation length we adopted here.

Let us consider the constraint on the string tension $G\mu$ from the B-mode polarizations.
To keep the constraint conservative, we assume $r=0$.
In this case,
the most stringent constraint from the B-mode polarizations comes from the signals observed by BICEP2 at $\ell \sim 320$ and by 
POLARBEAR at $\ell \sim700$.
Hence we find that in the delayed scaling scenario the constraint on $G\mu$ from the B-mode polarization
can also be relaxed.
Requiring that the string-induced B-mode signals with (without) the lensing effect 
do not exceed the data measured by BICEP2 and POLARBEAR,
we plot in the left (right) panel of Fig.~\ref{fig:gmulimit} each of these upper bounds 
on the string tension as a function of the initial correlation length.\footnote{Here we omit the 
third band of POLARBEAR at $\ell \sim 1500$ that gives a negative value.} 
We find the constraint 
imposed by the B-mode polarizations is comparable to or more severe than that by the temperature fluctuations, 
depending on the criterion of the condition.
Since the small-scale temperature fluctuations are usually dominated by the contributions from 
point sources, the Sunyaev-Zel'dovich effect and other secondary effects, the B-mode measurement would
be a complementary probe for strings with delayed scaling and also help to check the systematics 
in the derived constraint from the temperature fluctuations.
Moreover, the different features between the string-induced temperature fluctuations and polarizations
provide the different dependence on the string parameters. 
These different behaviors suggest that the combined analysis would be quite essential not only 
to obtain a tighter constraint on string parameters, but also to break the parameter degeneracies.
This is left for a future study.

\begin{figure}[htbp]
\centering{
\includegraphics[width = 0.45\textwidth]{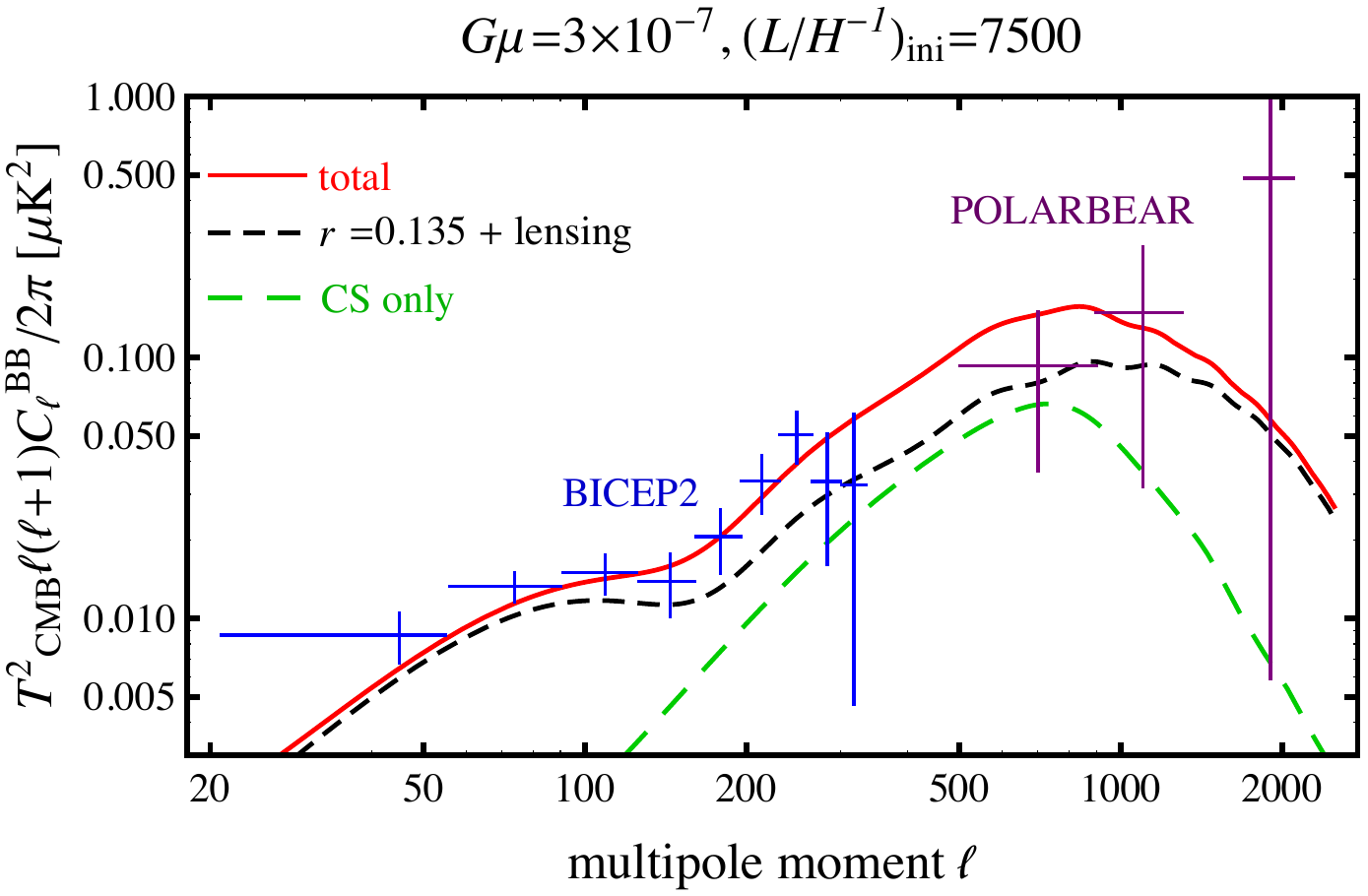}}
\caption{
B-mode polarization angular power spectra for gravitational lensing with
primordial gravitational waves ($r=0.135$, $n_t=0$, black short-dashed line) 
and contributions from cosmic strings with delayed scaling [$G\mu =3\times 10^{-7}$, $(L/H^{-1})_{\rm ini}=7500$, green long-dashed line].
If we combine them, the favorable value of $r$ would be smaller than 0.2.
}
\label{fig:fitting}
\end{figure}

Before closing this section, 
we attempt to match the string-induced B-mode spectrum to the BICEP2 and POLARBEAR data.
First we comment on the string-only model with gravitational lensing 
to compare the observational data.
As can be seen from the right panel of Fig.~\ref{fig:TTandBB},  
cosmic strings cannot create a small peak around $\ell\sim50$ and a small dip around $\ell\sim 150$, 
which are seen in the BICEP2 result,
though they can contribute to the higher multipoles.
Hence, even if we take into account the delayed scaling scenario, 
it is difficult to explain BICEP2 data only by cosmic strings 
since it cannot explain the BICEP2 result for both $\ell\lesssim 150$ and $\ell \gtrsim 150$ simultaneously, 
as well as POLARBEAR results.
Then, we consider the case in which both cosmic strings and primordial gravitational waves 
contribute to the B-mode signals.
Figure~\ref{fig:fitting} shows the angular power spectra of the B-mode polarization fluctuations 
induced by the cosmic strings with $G\mu=3\times 10^{-7}$ in the delayed scaling scenario 
[$(L/H^{-1})_{\rm ini}=7500$], 
as well as that induced by the primordial gravitational waves with $r=0.135$ and the gravitational lensing. 
We also plot the B-mode signals observed by BICEP2 and POLARBEAR.

\begin{figure}[htbp]
\centering{
\includegraphics[width = 0.45\textwidth]{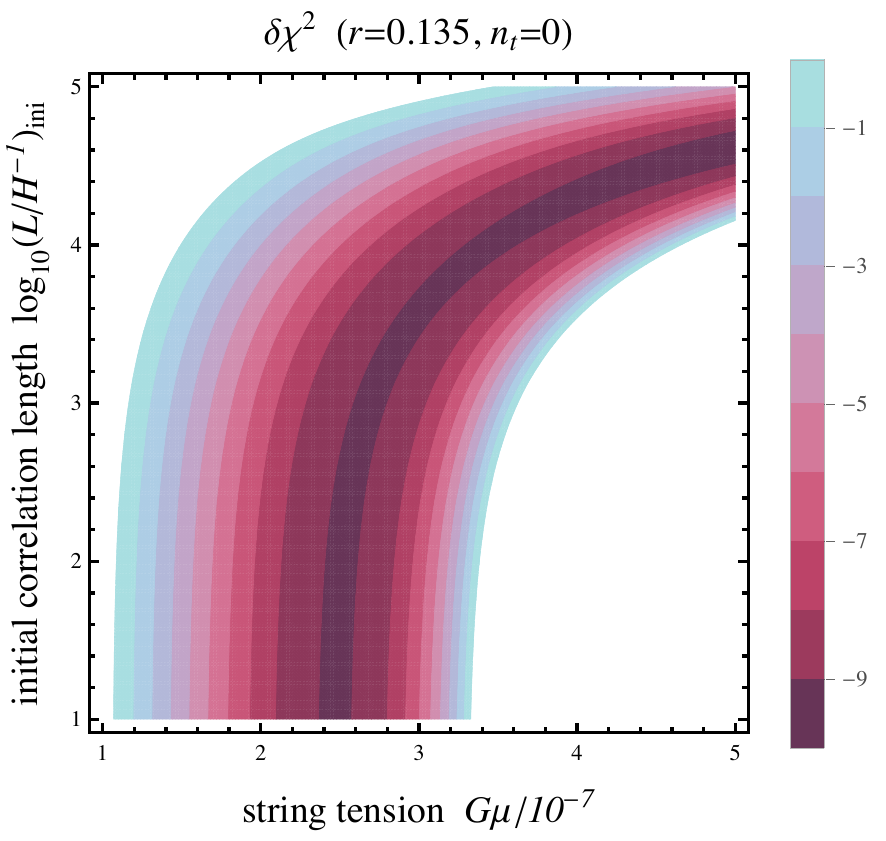}}
\caption{
The contour of the difference of $\chi^2$ between the prediction in taking $r=0.2$ without strings ($\chi^2\approx 15$)
and that of cosmic strings with delayed scaling added to that for $r=0.135$ based on BICEP2 data.
The value of $\chi^2$ gets worse in blank regions.
}
\label{chi2}
\end{figure}

We find this combination of three types of tensor perturbations significantly improves the fit to BICEP2 data.
As a demonstration, we show $\delta \chi^2$ as a function of $G\mu$ and $(L/H^{-1})_{\rm ini}$  in Fig.~\ref{chi2}, 
comparing the prediction of the delayed scaling scenario added to the primordial gravitational waves $r=0.135$ and 
the gravitational lensing with that of  the primordial gravitational waves $r=0.2$ and gravitational lensing 
without strings. 
It decreases as much as $\delta \chi^2 \approx -9$ with two additional parameters, 
$G\mu$ and $(L/H^{-1})_{\rm ini}$ as far as fitting to BICEP2 data is concerned.
We note here that compared to the conventional early scaling case,
the delayed scaling model predicts suppression of the power spectra at
high $\ell$ ($\ell \gtrsim 200$).
Therefore, we conclude that although the improvement of $\chi^2$ is
almost the same as the initially scaling case, which has been studied 
in Ref.~\cite{Lizarraga:2014eaa},
the detailed effects are different.
Moreover, we expect that the delayed scaling would be favored compared to
the conventional model if temperature data are included.
This is because in the delayed scaling scenario, a large enough string
tension to fit the B-mode data,
which is excluded in the conventional early scaling scenario, is still allowed
owing to the suppression of the temperature power spectra at
high $\ell$ as we have seen in the previous subsection.


\section{Conclusion and discussion \label{sec:4}}

In this paper we have studied the evolution of ``delayed scaling'' strings and 
its potential impacts on CMB measurements such as Planck, BICEP2 and POLARBEAR.
We have also discussed how the constraints on the parameters change.
The key is to consider the scenario in which cosmic strings are formed not after but
{\it during} inflation. 
Such strings have exponentially large separation due to
the dilution during the subsequent inflation and 
their evolution is quite different from that of strings
which enter the scaling regime at an earlier epoch.

We have traced typical evolution of the string network by solving 
the velocity-dependent one-scale model.
We have shown that if we take the relatively large correlation length at the initial time,
the correlation length decays as $a$ rather than $1/H$ at an earlier epoch
and it takes a few orders of redshift for the system to enter the scaling regime 
(Fig.~\ref{fig1}).

Based on the evolution of the network, we have calculated the angular power spectra
for the string-induced temperature anisotropies and B-mode polarizations.
We found that the large initial correlation length and the consequent delay of the entrance into
the scaling regime allow the decrease in the number of strings at an earlier epoch,
leading to the decay of the string signals mainly on higher multipoles 
(Fig.~\ref{fig:TTandBB}).
As a result, the delayed scaling scenario can relax the constraint on the string tension
from the measurements for both the temperature anisotropies and the B-mode polarizations.

We have further discussed the features of the B-mode signals produced by strings.
For the string-only model with the contribution of the gravitational lensing, it is 
difficult to explain both BICEP2 and POLARBEAR data fully 
since the shape of the spectrum from cosmic strings 
still lacks powers on lower multipoles as required by the BICEP2 results.
On the other hand, we have shown that delayed string contribution added to
the primordial gravitational waves with  smaller tensor-to-scalar ratio 
being consistent with Planck
improves the fit to the BICEP2 data,
although more data and its analysis are certainly required to draw final conclusion.

Throughout the paper, we have assumed several idealizations.
In particular, our results are based on 
the velocity-dependent one-scale model, 
though it is not obvious that this model 
characterizes the evolution of strings with delayed scaling.
The uncertainty may lead to significant changes in the observables.
Studies on the network model for the strings with delayed scaling should be further improved
for future applications to CMB measurements. 

\section*{Acknowledgments}
The authors thank D.~Figueroa, M.~Hindmarsh, and T.~Suyama for helpful discussions and comments. 
This work has been supported in part by the JSPS Postdoctoral Fellowships for Research Abroad (K.K.),
the Grant-in-Aid for JSPS Fellows No.~259800 (D.Y.), JSPS Grant-in-Aid for Scientific Research No.23340058 (J.Y.),
and JSPS Research Fellowships for Young Scientists (Y.M.). 

\appendix

\end{document}